\documentclass[preprint,preprintnumbers,showpacs,aps,prd,amssymb]{revtex4-1}

\usepackage{graphicx}
\usepackage{bm}
\usepackage{amsmath}

\def\Mhat{{\hat M}}

\def\nn{\nonumber}

\begin{document}
\title{On the quantization of the noncommutative geometry inspired black hole}
\author{Jong-Phil Lee}
\email{jplee@kias.re.kr}
\affiliation{Institute of Convergence Fundamental Studies, Seoul National University of Science and Technology, Seoul 139-931, Korea}
\affiliation{Division of Quantum Phases $\&$ Devices, School of Physics, Konkuk University, Seoul 143-701, Korea}

\begin{abstract}
Black holes in noncommutative geometry background are considered to be quantized in accordance with the holographic principle.
Incomplete gamma function involving the effective black hole mass is replaced by a discrete sum.
The mass spectrum as well as the temperature of black holes is presented.
The spectra are discrete ones but the shapes are very consistent with the known continuous results.
\end{abstract}

\maketitle
Understanding gravity in the language of quantum theory remains one of the most important issues in physics.
Black holes have provided a good test bed not only for gravity but for quantum mechanics also.
The Hawking radiation is the first fruit of combining gravity and quantum theory in black holes \cite{Hawking74,Hawking75}.
But there are still so many issues to be explained.
Among them is the final fate of the black hole through its evaporation via the Hawking radiation.
According to Hawking, black holes have temperature inversely proportional to their masses, so the black hole temperature tends to increase
indefinitely at the final stage of evaporation.
However it is widely believed that some kind of quantum rules would work out for taming the singular behavior of black holes.
In this work we adopt the noncommutative geometry background \cite{Witten, Seiberg}.
It is one of the most promising candidate for describing quantum nature of the spacetime.
There already have been many works on black holes in the noncommutative geometry 
\cite{Nicolini05,Nicolini06,Myung06, Ansoldi06,Banerjee08,Banerjee082,Arraut09,Banerjee09,Smailagic10, Mureika11}.
(For discussions in higher dimensions, see \cite{Rizzo06,Nicolini08,Nicolini11}.)
In this picture the gravitational source is not distributed as a delta function which is responsible for the normal Schwarzschild metric, 
but as a Gaussian with a dispersion of order $\sim \sqrt{\theta}$ where $\theta$ measures the noncommutativity of spacetime
\cite{Smailagic, Smailagic2}:
\begin{equation}
 \rho_{cont} (r)=\frac{M}{(4\pi\theta)^{3/2}}\exp (-r^2/4\theta)~,
\label{rhocont}
\end{equation}
where $M$ is the black hole mass parameter.
The Gaussian distribution seems quite plausible because in a noncommutative geometry there is a fundamental uncertainty in a small region
of order $\sim\theta$.
One of the results of this scheme is that there is a minimum mass (and consequently minimum horizon radius) 
below which the event horizon does not exist.
At this point the Hawking temperature vanishes and no singular behavior occurs.
\par
In this paper we focus on the fact that there is a minimum size of black holes in noncommutative geometry.
Inspired by this result we assume that the spacetime geometry inside every black hole is quantized 
in such a way that the surface area is quantized in units of the minimum area given by the minimum black hole radius.
As a consequence, every black hole in noncommutative geometry is provided with its surface area only in the form of multiples of fundamental area
defined by the minimum horizon radius.
In this sense we are adopting the holographic principle where all the relevant information about the black hole resides on its surface,
in accordance with the Bekenstein's entropy criterion \cite{Bekenstein72, Bekenstein73, Bekenstein74}.
In the literature, usually the noncommutative effects are encoded in the incomplete gamma function for the ``effective mass'' of the black hole,
which is given by the integral of the Gaussian density distribution over the black hole volume:
\begin{equation}
 m_{cont}(r)=\int_0^r 4\pi r'^2 \rho_\theta(r')dr'=\frac{2M}{\sqrt{\pi}}\gamma\left(\frac{3}{2},\frac{r^2}{4\theta}\right)~,
\label{mcont}
\end{equation}
where
\begin{equation}
 \gamma(a,x)=\int_0^x r^{a-1}e^{-t}dt~,
\label{gamma}
\end{equation}
is the incomplete gamma function.
But now that we are assuming the spacetime quantization, the integral resulting in the incomplete gamma function should be replaced by discrete sums.
This is the main point of current analysis.
The result is that black hole mass, in association with the event horizon radius, as well as the Hawking temperature, 
is quantized through simple algebraic functions.
\par
We first assume that the black hole area is {\em pixelated} in the sense that the area is quantized by some minimal or fundamental area.
The minimal area is closely related to the minimal horizon radius $r_0$ of the noncommutative black holes, as indicated in \cite{Nicolini06,Spallucci}.
In a compact form, the black hole surface area is assumed to be quantized as
\begin{equation}
 A_n=4\pi r_h^2=4\pi r_0^2 n~,
\end{equation}
where $r_h$ is the horizon radius and $n=1, 2, \cdots$.
A similar quantization appears in \cite{Spallucci}.
The quantization rule for $r_h$ is then 
\begin{equation}
 r_h=r_0\sqrt{n}~,
\label{rn}
\end{equation}
where $r_0\sim ({\rm a~few})\times\sqrt{\theta}$ is a small quantity of minimal length.
Strictly speaking, the quantization rule of Eq.\ (\ref{rn}) holds for the event horizon of black holes,
but here we simply assume that the same quantization holds for ordinary space.
All the relevant results of this work are based on these two assumptions.
First consider the mass density $\rho_\theta(r)$. 
Eq.\ (\ref{rhocont}) now should be changed as
\begin{equation}
 \rho_\theta(r)=\frac{M}{N_0}e^{-\frac{r^2}{4\theta}}~,
\label{rho}
\end{equation}
where $N_0$ is some normalization constant.
In ordinary noncommutative approach, $N_0$ is fixed to make the volume integral of $\rho_\theta(r)$ over the whole space equal to $M$.
For continuum case, 
\begin{equation}
N_0=\int_0^\infty e^{-r^2/4\theta}~4\pi r^2 dr=(4\pi\theta)^{3/2}~,
\end{equation}
as is given in Eq.\ (\ref{rhocont}).
But for a quantized spacetime (or if we consider an infinitely large black hole), the quantization rule of Eq.\ (\ref{rn}) must be applied. 
In this case the volume element to be summed is $4\pi(r_0\sqrt{n})^2\cdot r_0$
(we assume that $r_0$ is small enough so that the mass density does not change significantly for the range of $\Delta r\sim r_0$), 
and the continuous integral should be replaced by the discrete sum to give
\begin{eqnarray}
 N_0&=&\sum_{n=1}^\infty e^{-r_0^2 n/4\theta}~4\pi(r_0\sqrt{n})^2\cdot r_0\nn\\
&=&(4\pi r_0^3)\frac{e^\alpha}{(e^\alpha-1)^2}~,
\label{N0}
\end{eqnarray}
where $\alpha\equiv r_0^2/4\theta$.
The mass surrounded by a sphere of radius $r=r_0\sqrt{N}$, $m_\theta(r)$, is now given by
\begin{eqnarray}
 m_\theta(r)
&=&\sum_{n=1}^N 4\pi(r_0\sqrt{n})^2 r_0 ~\frac{M}{N_0}~e^{-r_0^2 n/4\theta}\nn\\
&=&M\left[1-(N+1)e^{-\alpha N}+Ne^{-\alpha(N+1)}\right]~.
\label{m}
\end{eqnarray}
\par
The spacetime metric is obtained by solving the Einstein equation with the mass distribution of Eq.\ (\ref{rho}).
For the case of continuum \cite{Nicolini06},
\begin{equation}
 ds^2=-f_{cont}(r)dt^2+\frac{dr^2}{f_{cont}(r)}+r^2d\Omega^2~,
\end{equation}
where
\begin{eqnarray}
 f_{cont}(r)&=&1-\frac{2m_{cont}(r)}{r}\nn\\
&=&1-\frac{4M}{r\sqrt{\pi}}\gamma\left(\frac{3}{2},\frac{r^2}{4\theta}\right)~.
\end{eqnarray}
Inspired by this result, the metric function $f(r)$ for the quantized spacetime should be
\begin{eqnarray}
f(r)&=&1-\frac{2m_\theta(r)}{r}\nn\\
&=&1-\frac{2M}{r_0\sqrt{N}}\left[1-(N+1)e^{-\alpha N}+Ne^{-\alpha(N+1)}\right]~.
\label{f}
\end{eqnarray}
Note that the continuous incomplete gamma function is replaced by the discrete sum.
\par
\begin{figure}
\includegraphics{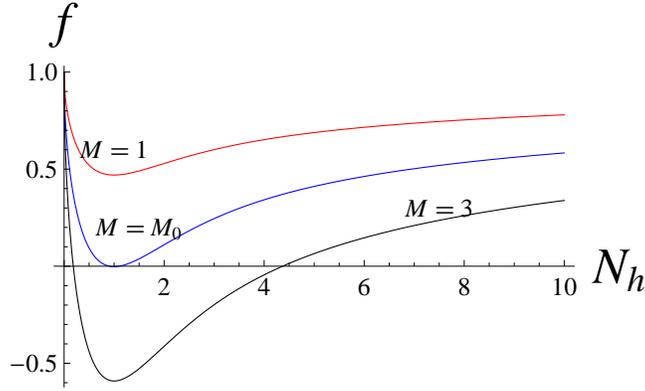}
\caption{Plots of $f(r)$ vs $N_h$ for various $M$.}
\label{fr}
\end{figure}
\begin{figure}
\includegraphics{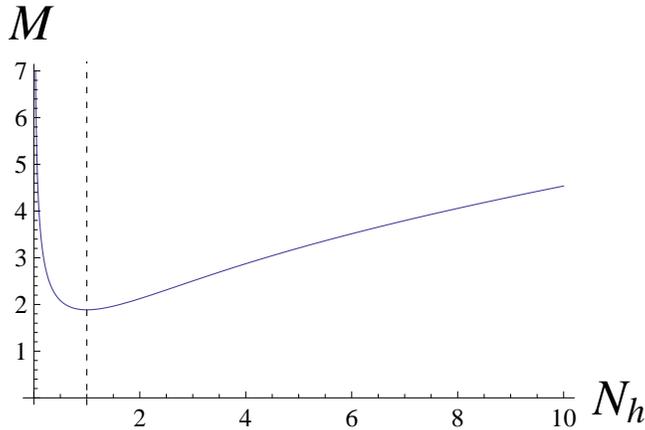}
\caption{Plots of $M$ (in units of $\sqrt{\theta}$) vs $N_h$.}
\label{MvsNh}
\end{figure}

The event horizon is determined by the condition $f(r_h)=0$, where $r_h$ is the horizon radius.
Letting $r_h=r_0\sqrt{N_h}$, one gets
\begin{equation}
 r_0\sqrt{N_h}=2M\left[1-(N_h+1)e^{-\alpha N_h}+N_h e^{-\alpha(N_h+1)}\right]~.
\label{horizon}
\end{equation}
As in the continuous case, there exists a minimum value of $M$ below which there are no horizons.
See Fig.\ \ref{fr}.
For the extremal case of $M=M_0$, there is only one horizon $r_h=r_0$.
At $r=r_0$, the metric function $f(r)$ satisfies $f(r_0)=f'(r_0)=0$, which is equivalent to $f(N=1)=f'(N=1)=0$ in our discrete case.
One finds that
\begin{equation}
 1-4\alpha e^{-\alpha}+(2\alpha-1)e^{-2\alpha}=0~,
\end{equation}
and the numerical solution is $\alpha=2.06$.
The minimal mass $M_0$ is then $M_0/\sqrt{\theta}=1.89$, which is consistent with that for the continuum case, $\sim 1.9$ \cite{Nicolini06}.
For $\alpha=2.06$, we have $r_0/\sqrt{\theta}=2.87$, which is slightly smaller than that for the continuum, $\sim 3.0$.
Fixing the value of $\alpha$, one can find the behavior of $M$ as a function of $N_h$ from Eq.\ (\ref{horizon}).
Figure \ref{MvsNh} shows the result.
Obviously $M$ gets its minimum value at $N_h=1$ (the vertical line) and increases as $N_h$, but not linearly as in the classical black holes.
Since $N_h$ is assumed to be integers from the quantization rule of Eq.\ (\ref{rn}), the black hole mass parameter $M$ is also quantized
according to Eq.\ (\ref{horizon}).
Explicitly, one can arrive at
\begin{equation}
 \frac{\sqrt{N_h}}{\Mhat_{N_h}}-\frac{\sqrt{N_h-1}}{\Mhat_{N_h-1}}=
N_h e^{-\alpha N_h}\frac{4}{\sqrt{\alpha}}\sinh^2\frac{\alpha}{2}~,
\label{Mn}
\end{equation}
where $\Mhat\equiv M/\sqrt{\theta}$.
Eq.\ (\ref{Mn}) is the main result of this paper.
Note that for sufficiently large $N_h$ (this is a classical limit) the right-hand-side of Eq.\ (\ref{Mn}) is negligible, 
and so $\Mhat_{N_h}\approx\Mhat_{N_h-1}$ for $N_h \gg 1$, showing a good correspondence to classical picture.
\par
\begin{figure}
\includegraphics{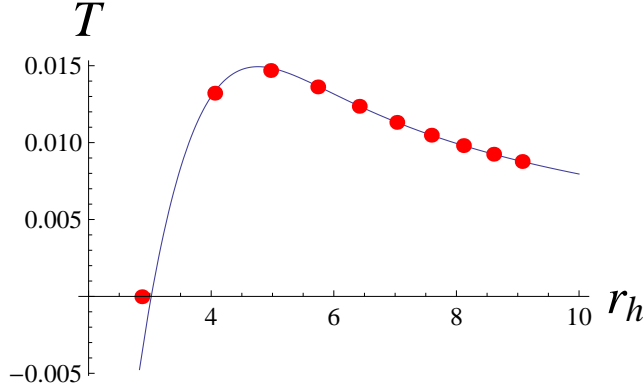}
\caption{Hawking temperature for continuum(solid line) and discrete(dots) spacetime in units of $1/\sqrt{\theta}$.}
\label{Th}
\end{figure}

Now it is quite straightforward to obtain the Hawking temperature.
The Hawking temperature is given by 
\begin{equation}
 T=\frac{1}{4\pi}\frac{d}{dr}f(r)\Bigg|_{r=r_h}
=\frac{1}{4\pi}\frac{2\sqrt{N}}{r_0}\frac{d}{dN}f(N)\Bigg|_{N=N_h}~,
\label{T}
\end{equation}
where the relation $r=r_0\sqrt{N}$ is used.
The result is
\begin{eqnarray}
 T&=&
\frac{1}{8\pi M}\frac{1+[N_h-1-2\alpha N_h(N_h+1)]e^{-\alpha N_h}+N_h(2\alpha N_h-1)e^{-\alpha(N_h+1)}}
{\left[1-(N_h +1)e^{-\alpha N_h}+N_h e^{-\alpha(N_h+1)} \right]^2}\nn\\
&=&
\frac{1}{4\pi}\frac{1}{r_0\sqrt{N_h}}\frac{1+[N_h-1-2\alpha N_h(N_h+1)]e^{-\alpha N_h}+N_h(2\alpha N_h-1)e^{-\alpha(N_h+1)}}
{1-(N_h +1)e^{-\alpha N_h}+N_h e^{-\alpha(N_h+1)}}~.
\label{TNh}
\end{eqnarray}
Figure\ \ref{Th} shows the Hawking temperature for continuum(solid line) and discrete(dots) spacetime.
It is quite impressive that the two results are very consistent with each other.
Note that $T(N_h=1)=0$, which is obvious from Eq.\ (\ref{T}) and the requirement $f(N=1)=f'(N=1)=0$ above.
The maximum temperature is $T(N_h=3)\simeq0.015/\sqrt{\theta}$, which is consistent with the continuum case \cite{Nicolini06}.
\par
It must be pointed out that the basic quantization rule is based on the holographic principle which asserts that the spacetime is pixelated.
The quantization rule of Eq.\ (\ref{rn}) is a direct consequence of the black-hole area quantization, which is originated from the Bekenstein's
area law.
For example, one could have imposed the quantization like $r_h=r_0\times n^k$ ($k\ne 1/2$) for some other reasons.
Then the whole results of current analysis look much different (especially the spectral patterns of $M$ or $T$) from those given here.
\par
In conclusion, we have provided with the quantization rule for the noncommutative geometry inspired black holes.
The rule is based on the holographic nature of black holes where their entropy is proportional to the surface area, 
and the fact that there is a minimum mass for the noncommutative black holes.
The quantization rule derived here is very specific with respect to the holographic principle; 
were it not for holography, the final form of the quantization rule might look much different from that presented in this paper.
In this sense, verification of our quantization rule is very important for checking whether the spacetime is really holographically quantized or not.
It will be also very interesting to extend the idea adopted in this work to noncommutative black holes in higher dimensions.
In that case it might be possible to see the quantized spectrum of black holes at the LHC.
\par
\begin{acknowledgments}
This work  was supported by  NRF grant funded by MEST(No. 2011-0029758).
\end{acknowledgments}


\begin{thebibliography}{99}
\bibitem{Hawking74}
S.~W.~Hawking,
  Nature {\bf 248}, 30 (1974).
\bibitem{Hawking75} 
  S.~W.~Hawking,
  Commun.\ Math.\ Phys.\  {\bf 43}, 199 (1975)
  [Erratum-ibid.\  {\bf 46}, 206 (1976)].
\bibitem{Witten}
E.~Witten,
  Nucl.\ Phys.\ B {\bf 460}, 335 (1996).
\bibitem{Seiberg}
 N.~Seiberg and E.~Witten,
  JHEP {\bf 9909}, 032 (1999).
\bibitem{Nicolini05} 
  P.~Nicolini,
  J.\ Phys.\ A A {\bf 38}, L631 (2005).
\bibitem{Nicolini06}
P.~Nicolini, A.~Smailagic and E.~Spallucci,
  Phys.\ Lett.\ B {\bf 632}, 547 (2006).
\bibitem{Myung06} 
  Y.~S.~Myung, Y.~-W.~Kim and Y.~-J.~Park,
  JHEP {\bf 0702}, 012 (2007).
\bibitem{Ansoldi06} 
  S.~Ansoldi, P.~Nicolini, A.~Smailagic and E.~Spallucci,
  Phys.\ Lett.\ B {\bf 645}, 261 (2007).
\bibitem{Banerjee08} 
  R.~Banerjee, B.~R.~Majhi and S.~Samanta,
  Phys.\ Rev.\ D {\bf 77}, 124035 (2008).
\bibitem{Banerjee082} 
  R.~Banerjee, B.~R.~Majhi and S.~K.~Modak,
  Class.\ Quant.\ Grav.\  {\bf 26}, 085010 (2009).
\bibitem{Arraut09} 
  I.~Arraut, D.~Batic and M.~Nowakowski,
  Class.\ Quant.\ Grav.\  {\bf 26}, 245006 (2009).
\bibitem{Banerjee09} 
  R.~Banerjee, S.~Gangopadhyay and S.~K.~Modak,
  Phys.\ Lett.\ B {\bf 686}, 181 (2010).
\bibitem{Smailagic10} 
  A.~Smailagic and E.~Spallucci,
  Phys.\ Lett.\ B {\bf 688}, 82 (2010).
\bibitem{Mureika11} 
  J.~R.~Mureika and P.~Nicolini,
  Phys.\ Rev.\ D {\bf 84}, 044020 (2011).
\bibitem{Rizzo06} 
  T.~G.~Rizzo,
  JHEP {\bf 0609}, 021 (2006).
\bibitem{Nicolini08} 
  P.~Nicolini,
  Int.\ J.\ Mod.\ Phys.\ A {\bf 24}, 1229 (2009).
\bibitem{Nicolini11} 
  P.~Nicolini and E.~Winstanley,
  JHEP {\bf 1111}, 075 (2011).
\bibitem{Smailagic} 
  A.~Smailagic and E.~Spallucci,
  J.\ Phys.\ A  {\bf 36}, L467 (2003).
\bibitem{Smailagic2} 
  A.~Smailagic and E.~Spallucci,
  J.\ Phys.\ A  {\bf 36}, L517 (2003).
\bibitem{Bekenstein72} 
  J.~D.~Bekenstein,
  Lett.\ Nuovo Cim.\  {\bf 4}, 737 (1972).
\bibitem{Bekenstein73} 
  J.~D.~Bekenstein,
  Phys.\ Rev.\ D {\bf 7}, 2333 (1973).
\bibitem{Bekenstein74} 
  J.~D.~Bekenstein,
  Phys.\ Rev.\ D {\bf 9}, 3292 (1974).
\bibitem{Spallucci} 
  E.~Spallucci and A.~Smailagic,
  Phys.\ Lett.\ B {\bf 709}, 266 (2012).
\end{thebibliography}
\end{document}